# Mixed Reality using Illumination-aware Gradient Mixing in Surgical Telepresence: Enhanced Multi-layer Visualization

Nirakar Puri[1], Abeer Alsadoon[1,2,3,4*], P.W.C. Prasad[1], Nada Alsalami[5], Tarik A. Rashid[6]

[1]School of Computing and Mathematics, Charles Sturt University (CSU), Wagga Wagga, Australia
[2]School of Computer Data and Mathematical Sciences, University of Western Sydney (UWS), Sydney, Australia
[3]Kent Institute Australia, Sydney, Australia
[4]Asia Pacific International College (APIC), Sydney, Australia
[5]Computer Science Department, Worcester State University, MA, USA
[6] Computer Science and Engineering, University of Kurdistan Hewler, Erbil, KRG, Iraq.

Abeer Alsadoon[1*]
* Corresponding author. A/Prof (Dr) Abeer Alsadoon, [1]School of Computing and Mathematics, Charles Sturt University, Sydney Campus, Australia.
Email: alsadoon.abeer@gmail.com , Phone +61 413971627

**Abstract:**
***Background and aim:*** Surgical telepresence using augmented perception has been applied, but mixed reality is still being researched and is only theoretical. The aim of this work is to propose a solution to improve the visualization in the final merged video by producing globally consistent videos when the intensity of illumination in the input source and target video varies.
***Methodology:*** The proposed system uses an enhanced multi-layer visualization with illumination-aware gradient mixing using Illumination Aware Video Composition algorithm. Particle Swarm Optimization Algorithm is used to find the best sample pair from foreground and background region and image pixel correlation to estimate the alpha matte. Particle Swarm Optimization algorithm helps to get the original colour and depth of the unknown pixel in the unknown region. ***Result:*** Our results showed improved accuracy caused by reducing the Mean squared Error for selecting the best sample pair for unknown region in 10 each sample for bowel, jaw and breast. The amount of this reduction is 16.48% from the state of art system. As a result, the visibility accuracy is improved from 89.4 to 97.7% which helped to clear the hand vision even in the difference of light. ***Conclusion:*** Illumination effect and alpha pixel correlation improves the visualization accuracy and produces a globally consistent composition results and maintains the temporal coherency when compositing two videos with high and inverse illumination effect. In addition, this paper provides a solution for selecting the best sampling pair for the unknown region to obtain the original colour and depth.

**Keywords:**
Augmented reality, Virtual reality, Mixed reality, Surgical telepresence, Visualization, Illumination-Aware, Image pixel correlation, particle swarm optimization

## 1. Introduction:

Expert surgeons are not available globally at all times, and some complex surgical conditions need a real-time assistance from another expert surgeon, either virtually or physically by attending the local site. Nevertheless, on-site surgery attendance for the expert surgeon is time-consuming and may not be possible all the time [1]. In addition, the patient's Computed Tomography (CT) scan reports should be generated in the preparation stage of the operation and shared with the pre-planning experts. This process is expensive and time-consuming and was replaced by video-guided monitoring system, which allowed virtual communication between the local site (surgical site) and the remote site (expert surgeon site) [2].

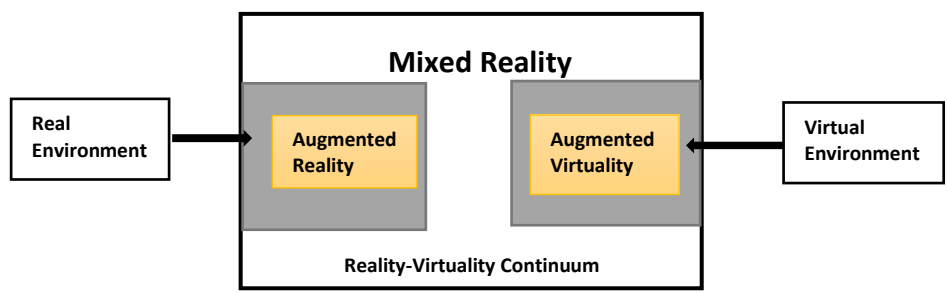

Figure 1: Reality- virtuality (RV) continuum.

Mixed Reality (MR) is introduced to contribute the remote collaboration that combines the virtual world and the real world, and this concept is called Reality-Virtuality continuum as shown in Figure 1[1]. The augmented video is created on the surgery site. The experts on the remote site interact with the augmented video and can guide the rural surgeon using hand gesture that will be recorded by camera. The MR view is created from merging augmented and virtual videos and is shared to both sites. MR has allowed expert surgeons to assist in real live surgery at local sites directly from their virtual hand [3]. MR is promising surgical guidance technology [4]. It can boost the sensory experience of the surgeon by combining imaging modalities with real objects and enable the patient to interpret depth and spatial interactions in the realistic world. Unlike Augmented Reality (AR) systems that can only overlay the visual image on the real object, MR system can provide a fully interactive holographic environment in which the user can interact with virtual object while the objects respond to complex interaction including physically-based contact modelling, deformation and so on [5][6].

The remaining sections of this paper is structured as a Literature Review section that discusses the latest solutions and describes the current best solution, i.e. the Enhanced Multi-layer mean value cloning Algorithm [1] (EMMVC). The proposed system is explained in sections 3. The Result (section 4) provides the tested sample and visual data whereas the discussion (section 5) provides the discussion of the results. Finally, the research concludes by providing a conclusion and future works to be done in section 6.

## 2. Literature Review:

The literature review examines and evaluates the existing research and systems and find the opportunities to improve them. The evaluation offers an understanding of the various methods, procedures, and instruments used in the field of study. This section provides an evaluation and an analysis for various published papers related to our work. The aim of this work is to combine the augmented video (generated at the local site) with the surgeon's hand's virtual video. This combination produces a high level of accuracy MR view and is sent to both sites. The combined video can restore the patient's augmented image pixels which are occluded by the hands and instruments of the surgeon, and virtual image pixels which are influenced by the illumination intensity difference in operation theatre. For better understanding, we organized the studied papers into five subsections according to their main method. These subsections are: Capturing Motion of Hand, Illumination and colour correction Method, Remote collaboration Systems using Mixed Reality, Image Restoration, and Enhanced Multi-layer mean value cloning Algorithm.

**2.1** Capturing Motion of Hand

Shakya et al.[2] enhanced Continuously Adaptive Mean SHIFT (CAMSHIFT) algorithm. CAMSHIFT algorithm helps to detect the high-speed motion of the expert surgeon hand during the surgery, and to remove the occlusion. The research solved the occlusion problem by using two Red-Green-Blue-Depth (RGBD) cameras to get the initial alpha values and to get a wide angle of the hand. This solution conducted research on 10 breast samples using extended CAMSHIFT algorithm. The results reduced the average accuracy error of the video frame image overlay from 1.44mm to 1.28mm with the average processing time enhanced from 76 second compared to the state of art method. Venkata et al. [1] used the trimap contour flow algorithm and tripmap propagation algorithm to capture the motion of the hand. Nevertheless, Venkata et al. [1] and Shakya et al. [2] have not considered any effect of illumination intensity when merging the video frames [7]. There will be a different light intensity in the operation theatre and the expert surgeon hand capturing area. As a result, the merged video will have the light effect.

**2.2** Illumination and colour correction Method

Venkata et al. [1] proposed a solution to enhance the multi-layer mean value cloning algorithm to downgrade overlay error, improve processing time and improve the visibility accuracy by decreasing the occlusion, smudging and discoloration pattern. The research used EMMVC algorithm, trimap generation and alpha matting technique. This solution conducted research on 10 soft and 10 hard tissue image frames. As a result, the accuracy in term of overlay error was improved from 1.3mm to 0.9 mm and the visualization accuracy was also improved from 98.4% to 99.1%. The processing time was minimized from 11 seconds to 10 seconds for 50 frames. However, this solution has not used the correlation for alpha matting and thus suffer from artefacts around the merged region. The global alpha matting approach used in this work was unable to find the best pixel sample for the unknown region [8].

Wang et al. [7] and Pluhacek et al. [9] extended the mean-value cloning by smoothing the discrepancies between the source frame and the target frame and produced a globally consistent composition results. Also, these works maintain the temporal coherency when compositing the two videos with high and inverse illumination effect. The two research offered a solution to the illumination problem by optimizing the interpolation for image cloning, and the spatial-temporal consistent blending boundary. Also, they offered illumination guided gradients mixing. As a result, the processing time was 24.2 seconds in which optical flow consumes 22.7 seconds. The user supervision is not required so often, and the quality of the final composite video is very high and does not suffer from any discoloration and illumination intensity varying. Henry et al.[10] used the automatic trimap generation to solve the colour correction.

Yan et al. [8] and Archer et al. [11] proposed image pixel correlation sampling method to reduce the artefacts that occur when alpha matte in the image. Their solution used the artificial immune network (aiNet) to collect the sample set and applied Particle Swarm Optimization (PSO) algorithm to select the best sample, and finally integrated the pixel correlation to smoothen the pixel of unknown regions. The accuracy is 0.81 measured in term of Mean Squared Error (MES). It has been improved compared to the current methods.

## 2.3 Remote collaboration Systems using Mixed Reality

Fang et al. [3] proposed a new remote collaborative platform for remote 3D communication using Head Mounted Display (HDM) to support more natural and intuitive interaction. Fang et al. [3] improved the co-presence awareness and collaborative efficiency. This solution used Basler camera in the local workers site to capture the scene and transfer it to the remote site as augmented video in 3D. Also, this method superimposes the hand gesture of the remote helper hands in the 3D video of the local site and sends it back as a guideline for the local person seeking help. The research used the Dynamic link library algorithm for this task and provided a consistency of 0.852 measured as a Cronbach's alpha, and an improvement of processing time by 36.43%. The consistency and processing time are considered acceptable to solve the remote collaboration for teaching and guiding.

Huang et al. [12] provided a specific type of remote-collaboration system where a remote helper guides a local worker using an audio communication and hand gestures to perform a repair or a maintenance task. This process gives more advantage over the current 2D based collaboration systems. The offered solution used the 3D augmented video of the local workspace and fusing that video to the remote helpers' hands gesture to create a real time MR for the helper. The Standard deviation of Immersion and the overall usability were 0.99 and 0.96, while for the old system they were 0.94 and 0.95, respectively. As a result, the usability in term of immersion effect is accepted.

Si et al. [5], Wang et al. [13], and De lima et al. [14] proposed new systems for remote collaboration and guidance in abdominal surgery for needle placement using MR. The Authors had proposed three components namely heterogeneous anatomy construction, virtual-real spatial information visualization registration, and automatic registration. These systems improved the reconstruction of an image frame when creating the MR view. Compared to the traditional CT scan guided method, these systems show an improvement in the operation time and in precision [15].

## 2.4 Image Restoration

Basnet et al. [16] and Murugesan et al. [17] worked on image restoration for merging the images. Basnet et al. [16] reduced the processing time by using optical camera and enhanced the tracking and learning detection. The research used Iterative Closest Point (ICP) for image registration, and this improved the quality of the live video optical camera and the CT scan image with the live video. Basnet et al. [16] and Murugesan et al. [17] used the noise removal technique to improve the quality of live video frames. Also, they used the occlusion removal method which gave image overlay to 0.23~0.35 mm. Processing speed of 8~12 frames per second was also achieved.

Zhang et al. [18] and Oyekan et al. [19] improved the image restoration method after occlusion removing to have more efficient restoration for the occlusion region with significant textured structure. These papers solved the problem by using a spinning parallelogram operator algorithm to find the pixel in the damaged area for image restoration. Thus, it can effectively remove occlusion object and restore the damaged part with the object being 400mm in front of the original image. This work operated successfully on the textured structured image [20].

The aim of our research is to propose a new system that improve the accuracy by reducing the visualization error. Our proposed solution also finds the best value for an unknown pixel in an unknown region during the merging of image frames. Pixel correlation is used to find the best value of the pixel. The best current solution in this area of research is presented in section 2.5.

## 2.5 Enhanced Multi-layer mean value cloning Algorithm

The review section presents the studies done about the remote collaboration in surgery and the merging techniques for generating the MR video from the remote site augmented video and the local expert site virtual images. Based on these studies, the key aspect considered in our work are accuracy, processing time, depth perception, noise, occlusion, visualization, spatial-temporal consistency, and the movement of objects.

Venkata et al. **[1]** enhanced the multi-layer mean value cloning algorithm to reduce overlay error, improve processing time and improve the visibility accuracy by decreasing the occlusion, smudging and discoloration pattern. Venkata et al. **[1]** used EMMVC algorithm, trimap generation and alpha matting technique. The advantage of trimap generation is that it helps to achieve accurate objects from the image and identify the foreground, the background, and unknown regions clearly. The author has used trimap propagation and contour flow method for detecting the movement of the object which is very important in remote collaboration of surgery. This paper has worked toward the enhancement of the existing mean value cloning algorithm by adding more techniques to optimize the final outcome and reduce the processing time and remove the artefacts by applying alpha trimap generation, adding interpolation constraint coefficient (k) to modulate the color variation and multi-layer visualization technique to adjust the transparency value to make desired layer transparent **[1]**. This solution conducted research on different soft (breast) and hard (jaw) tissue image frames. As a result, the accuracy in term of overlay error has been improved from 1.3mm to 0.9 mm. The visualization accuracy improved from 98.4% to 99.1% and the processing time from 11 seconds to 10 seconds for 50 frames. After studying and analysing all research presented in this section, we selected this research, Venkata et al. [1] as the first best solution. All there researches The block diagram shown below in Figure 2 depicts the features and the limitations of the best current solution in blue and red dotted rectangles, respectively.

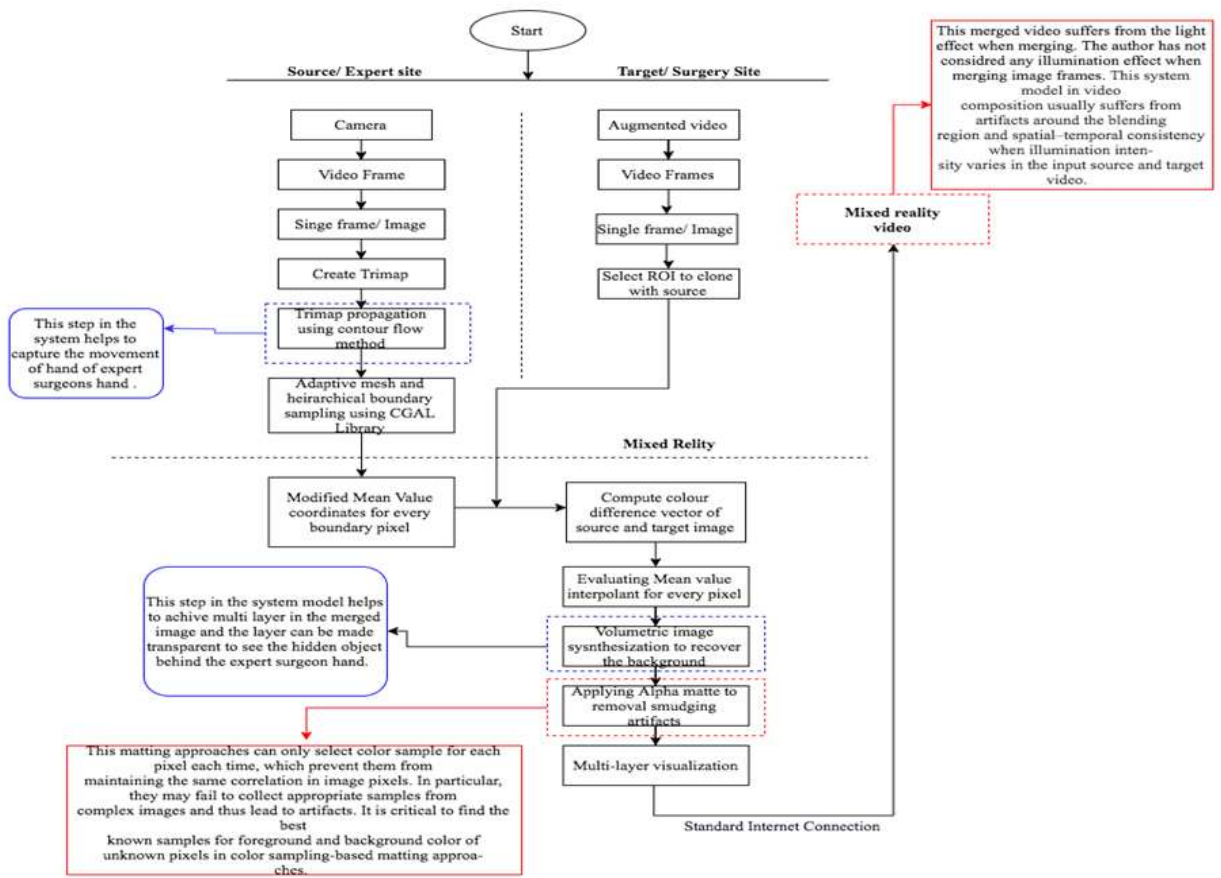

Figure 2: Enhanced Multi-layer mean value cloning Algorithm Solution Block Diagram **[1]**
[The. blue rectangle shows the good feature, and the red rectangles shows the limitation.]

Table 1: Pseudocode of the Enhanced Multi-layer mean value cloning system model.

---

**Algorithm: Enhanced Multi-layer mean value cloning Algorithm**
**Input:** boundary pixels' list δ b, boundary inner pixels' list $b_{in}$, trimap with three regions source image $I_s$ and target image colour vector $I_t$
**Output:** Merged video without artefacts.

**BEGIN**

**Step 1**: Generate the trimap for current and previous frames.
**Step 2**: If frame 0! = frame N then Calculate the contour flow between the two trimap using the equation:
$$I_{cf}(p) = (x_{current}, y_{current})$$
**Step 3**: Calculate the mean value coordinates of every inner pixel of generated source trimap generated in step 2 using this equation:
$$\lambda_i(p) = \frac{W_i}{\sum_{j=0}^{m-1} W_j} \quad i = 0, 1, \ldots\ldots m-1$$
**Step 4**: Repeat steps 2 and 3 and produces the cloned image for every image.
**Step 5**: Discoloured artefacts are removed in the cloned image using the interpolation constraint coefficient. k = 0 to 255, is considered as RGB channels. It can obtain by equation:
$$I_c(p) = \begin{cases} I_s(p) + k * r(p) & \text{for frame 0} \\ I_{cf}(p) + k * r(p) & \text{for frames 1, 2 \ldots N} \end{cases}$$
**Step 6**: Alpha matte is applied to smooth and remove the smudging artefacts. It can be given as equation:
$$I_c = \alpha I_c(p) + (1 - \alpha) I_t(p)$$
**Step 7**: A multi layered merged video is produced by using volumetric image synthesisation to recover the occlusion. It can get by equation:
$$I_{layers}(p) = \alpha I_c(p) + \beta I_c(p)$$

**END**

---

This solution consists of three stage: Remote expert surgeon site (source site), Local surgeon site (target) and composition site as shown in Figure 2. The three stages mentioned above are described as follow:

*Local Surgery Site (Target):* Augmented video based on the patient information and image is generated from the original generated video. The video frames are extracted, and those frames are sent to the remote expert surgeon site with the help of the internet so that the remote surgeon can perform a virtual surgery.

*Remote expert site (Source):* The expert Surgeon's hand action and the guidelines are recorded by the Red Blue Green Depth (RBGD) camera. From the recorded video, frames are extracted to be merged with patient video frame. From the two extracted frames, extracted from remote and local site, two consecutive extracted frames and two trimap's are generated respectively to detect the movement of expertise hand between the two frames. A trimap propagation technique is used to find the contour on the frames for which a user must specify the foreground, the background and the left one as unknown region. As a result, update the contour flow of the current frame to the next stage.

*MR (Video composition):* From the obtained trimap, an adaptive mesh is constructed using CGAL library. To reduce the processing time some inner pixels of the expert surgeon hand are ignored for the next stage based on the hierarchical boundary sample [1]. After this step, mean value are calculated for every pixel in the refined boundary region by using modified mean value algorithm. To remove the discoloration artefacts, an interpolation constraint coefficient k is introduced which maintains and manages the colour variations on both images. Temporal consistency is a key to achieve seamless video consistency and to gain it an alpha matte is generated that removes the smudging artefacts. Venkata et al. [1] used the volumetric image synthetization to generate a multi layers in the combined image which helps make the hand of the surgeon transparent. Finally, the refined and the merged reality video are sent to the local surgeon site.

*Limitation:* This solution has not considered any illumination effect when merging the source image frame and target image frame. Video composition usually suffers from artefacts around the blending region and spatial– temporal consistency when illumination intensity varies between the input source and the target video. When there is a light intensity difference between the target and the source object there will be a chance that the image with high illumination intensity will make other section invisible. This will confuse the surgeon and the visibility will be degraded very much. In addition, the matting approach can only select colour sample for each pixel each time, which prevent them from maintaining the same correlation in image pixels. In particular, they may fail to collect appropriate samples from complex images and thus lead to artefacts. It is critical to find the best-known samples for the foreground and the background colour of unknown pixels in the colour sampling-based matting approach. Figure 3 present the flow diagram of Enhanced Mean Value Cloning.

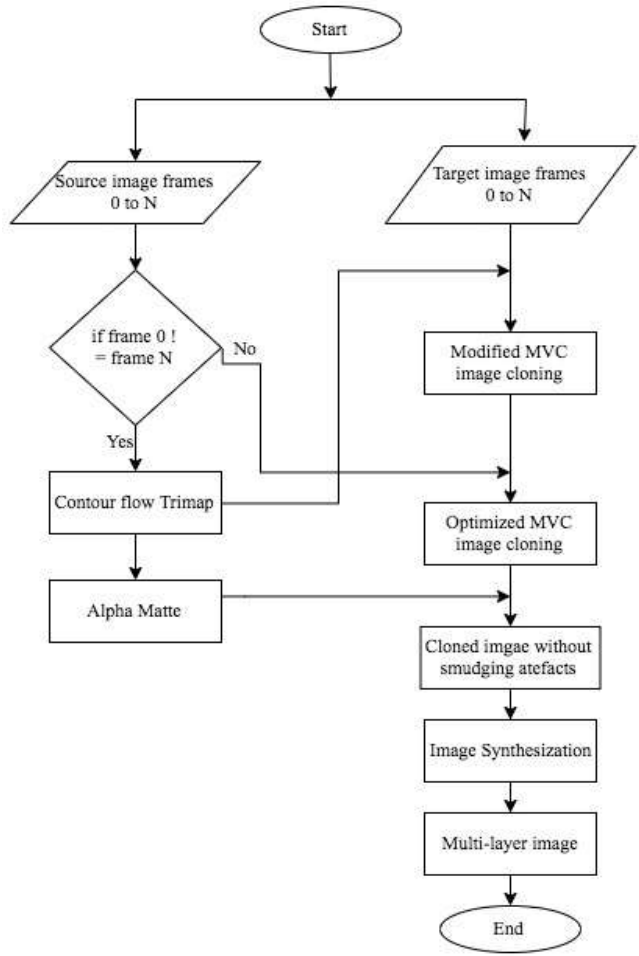

Figure 3: The flow diagram of Enhanced Mean Value Cloning.

A multi layered merged video is produced by using volumetric image synthesisation to recover the occlusion. It can get by the equation of Image multi-layer composition which is given as in Equation (1):

$$I_{layers}(p) = \alpha * HI_c(p) + \beta * I_{hc}(p) \qquad (1)$$

Where,
α, β are the blending parameters, where, α+β=1.
$HI_c(p)$ is the modified contour flow trimap for a given source, by using equation 2.
$I_{hc}(p)$ is the composited image and considered as the target image for multi-layer visualization. It is calculated by using Equation 3:

$$HI_c(p) = \begin{cases} I_s(p) + k * r(p) \: for \: frame \: 0 \\ I_{cf}(p) + k * Hr(p) \: for \: frame \: 1,2 \ldots \ldots N \end{cases} \qquad (2)$$

Where,
p is the inner pixel.
$I_s$ is RGB colour vector of the source image.
K is the interpolation constraint coefficient.
$I_{cf}$ is the motion of the source image trimap.
$r(p)$ is the interpolation vector.
$Hr(p)$ is the interpolation vector for the frame greater than zero and calculated using equation 6.

Alpha matte is applied to remove the smudging artefacts and generated based on the trimap. For given image Is and It as source and target images, image composition using alpha matting is to remove smudging and discolour artefacts is given by Equation 3.

$$I_{hc} = \alpha I_s(p) + (1-\alpha)I_t(p) \qquad (3)$$

Where,
p is inner pixel.
$I_{hc}$ is final composite image.
$I_s$ is the source image.
$I_t$ is target image.
α is alpha channel or transparency i.e. [0,1].
$I_c$ is final composite image with gradient mixing to remove illumination difference.

Depending on the above analysis, EMMVC [1] is selected as the best solution. It is selected based on the accuracy and the processing time of the implemented merging algorithm. EMMVC uses the trimap propagation with contour flow that gives the movement of the object by matching the previous and the current trimap. Multi-layer visualization is also used to remove the occlusion caused by the source image and its artefacts. Thus, EMMVC algorithm helps to merge the videos in optimized way by considering the facts about object movement, and occlusion removal based on multi-layer visualization. Figure 2 show the three phases of the EMMVC solution [1], and Table 1 shows the pseudocode of the EMMVC model [1].

## 3. Proposed System

Our aim is to propose a new solution to improve the visualization accuracy in the final merged video by producing globally consistent video when illumination intensity varies in the input source and target video. Our proposed system considered the method proposed by Venkata et al. [1] (i.e. EMMVC) the first-best solution. In addition, our system considered Illumination-Guided video composition via gradient consistency optimization suggested by Wang et al. [7] as the second-best solution. The illumination-guided video composition method helps to produce a globally consistent composition result even when the illumination intensity varies in the input source and the target video. This is a new feature adapted from the second-best solution. The composition technique can solve the illumination guided technique and thus helps the surgeons to visualize clearly without confusion even when there is a light difference in the image frames. Furthermore, Yan et al. [8] implemented the alpha matting with image pixel correlation for getting the unknown pixel in the colour-based matting approaches to remove the smudging artefacts. The EMMVC approach [1] is interpolation and can only select colour sample for each pixel each time, which prevent maintaining the same correlation in the image pixels. The EMMVC solution [1] failed to collect appropriate sample in a complex image and thus leads to artefacts. The correlation-based sampling method helps to define and select the best sample for unknown pixels, so that the true property of the pixel will not be lost. From the analysis and evaluation of EMMVC method proposed by Venkata et al. **[1]** ( given in section 2.5), we have concluded that there are two main limitations in this method. First, it does not consider any effect of light in source and target image frame when merging the video. Second, it does not consider neighbour pixel value when finding the image pixel for unknown pixel so that it has smudging artefacts in blending regions. Our proposed solution solved the first limitation by using illumination aware grading mixing when the light intensity varies in source and target video. Furthermore, the proposed method solved the second limitation by using image pixel correlation to find the best sample in the sample collection for unknown pixel. In the section, the proposed algorithm is explained in detail.

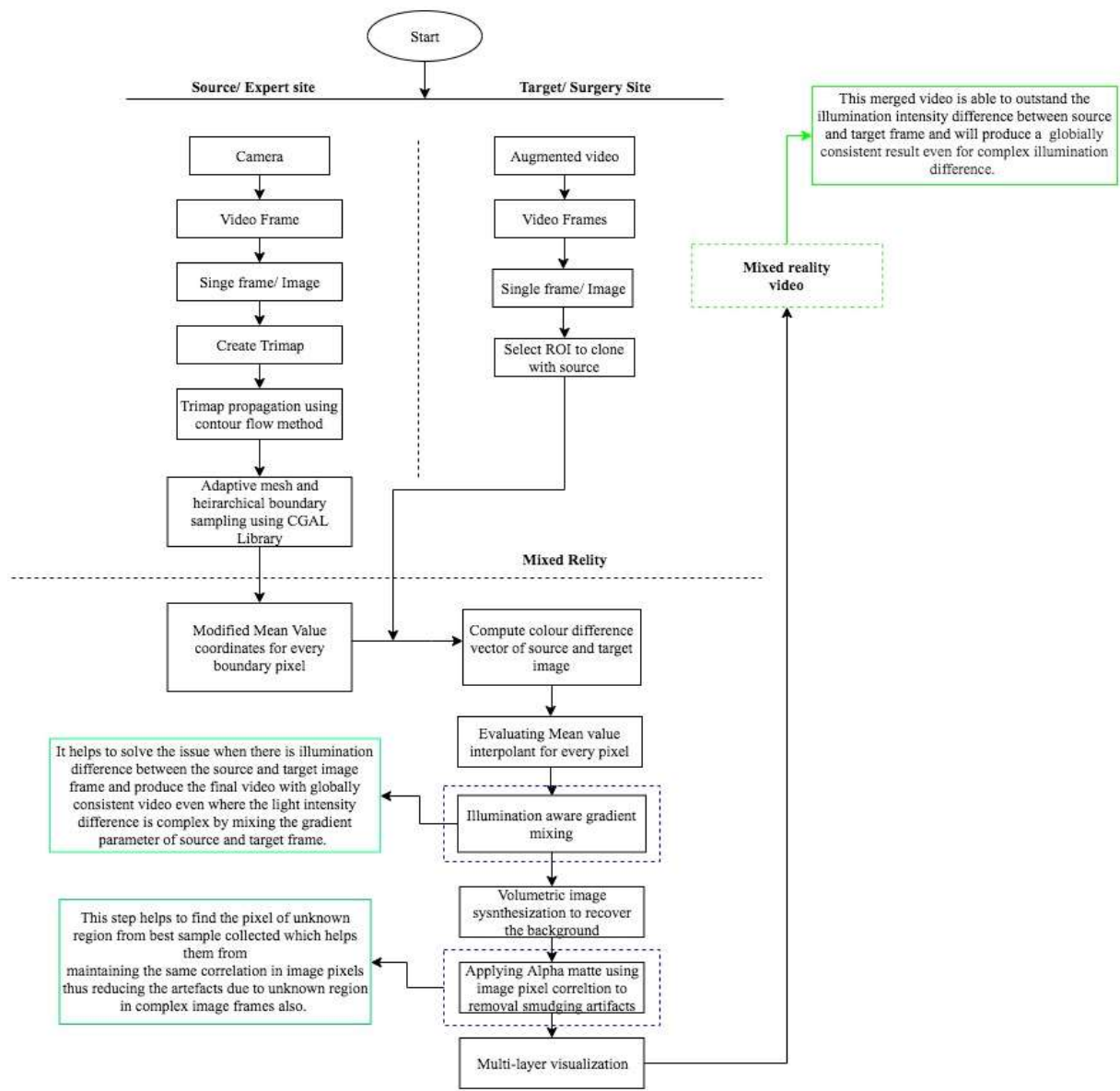

Figure 4: The Proposed solution Diagram
[The green rectangles refer to the modified parts in our proposed system]

The proposed system has three stages namely: local (surgery) site process stage, remote expert surgeon site process stage and the third one is video merging process stage where the merging of two videos will take place.

***Local Surgery Site (Target):*** The augmented video is generated based on the patient information and the generated image. From the original generated video, the video frames are extracted, and those frames are sent to the remote expert surgeon site so that the remote surgeon can perform a virtual surgery and the video composition can be done.

***Remote expert site (Source):*** In the remote site the augmented video is received from the first stage. Expert Surgeon's hand action and guidelines are recorded by the RBGD camera. From the recorded video, frames are extracted to be merged with the patient video frame. From the two extracted frames, two consecutive extracted frames and two trimap's are generated respectively to detect the movement of the expertise hand between the two frames [1]. A trimap propagation technique is used to find the contour on the frames for which a user has to specify the foreground, the background and the unknown region, and update the contour flow of the current frame to the next stage.

***MR (Video Composition):*** From the obtained trimap, an adaptive mesh is constructed using CGAL library. To reduce the processing time some inner pixels of the expert surgeon hand are ignored for the next stage based on the hierarchical boundary sample **[1]**. After this step, mean value are calculated for each pixel in the refined boundary region by using modified mean value algorithm. Then, Illumination guided gradient mixing is applied using the gradient mixing parameter**.** Temporal consistency is a key to achieve the seamless video consistency. An alpha matte is generated using image pixel correlation to find the best sample for the unknown region. PSO algorithm is applied to select the best sample from the collected sample set by artificial immune network. This helps to remove the smudging artefacts and the discoloration in the complex image frames [7]. Table 2 shows the pseudocode for the proposed system, while Figure 4 and Figure 5 show the diagram and the flowchart of the purposed system respectively.

### 3.1 Proposed Equation

The gradient mixing parameter is used to solve the problem of illumination difference between the source and the target videos. For given image Cs and Ct as source and target images, to remove illumination effect between image we must add gradient mixing parameters and is given by Equation 4 [7]. The calculation uses Equation 5 to find out the difference in the gradient between the source and the target image frame.

$$a_i = \frac{||G_i^s||^2}{||G_i^s||^2 + ||G_i^t||^2} + (C_t - C_s) \tag{4}$$

Where,
G denotes the mixing gradients parameter.
$G^s$ denotes the gradient parameter of the source image frame.
$G^t$ denotes the gradient parameter of the target image frame.
$a_i$ is the mixing weight factor at pixel i.
$C_t$ and $C_s$ are the differences of discolouration of the current and the next source frame.
The value of G can be calculated from Equation 5 [7]:

$$G = b \circ G^s + [(1,1) - b] \circ G^t \tag{5}$$

Where,
b = 1...k
$G^s$ denotes the gradient parameter of the source image frame.
$G^t$ denotes the gradient parameter of the target image frame.
∘ is the symbol of Hadamard product of the matrix.

The cost function for $a_i$ is calculated by modifying Equation 4 to reconsider the colour difference. Equation 6 calculates the gradient mixing parameter based on the gradient of the course and the target image frame. Equation 6 helps to determine the difference in the source and the target light intensity and will give the appropriate mixing parameter value that is used in our solution to make it globally consistent. This proposed modified equation will help to remove the illumination effect when merging two frames.

$$\mathrm{M}a_i = \frac{||G_i^s||^2}{||G_i^s||^2 + ||G_i^t||^2} \tag{6}$$

Where,
$\mathrm{M}a_i$ is $the\ modified\ illumination\ gradint\ mixing\ parameter$.
G denotes the mixing gradient parameter.
Gs denotes the gradient parameter of the source image frame.
Gt denotes the gradient parameter of the target image frame.

As a result, Equation 2 is modified to calculate the gradient mixing parameter and solve the illumination intensity difference limitation. The front part calculates the gradient mixing parameter and the remaining helps to calculate the colour difference, as given in Equation 7:

$$MHI_c(p) = \begin{cases} Ma_i + I_s(p) + k * r(p) \text{ for frame } 0 \\ Ma_i + I_{cf}(p) + k * Hr(p) \text{ for frame } 1,2 \ldots \ldots N \end{cases} \quad (7)$$

Where,
$M_{ai}$ is the illumination gradient mixing parameter.
p is the inner pixel.
$I_s$ is RGB colour vector of the source image.
K is the interpolation constraint coefficient.
$I_{cf}$ is the motion of the source image trimap.
$r(p)$ is the interpolation vector.
$Hr(p)$ is the interpolation vector for the frame greater than zero.

To solve the pixel by pixel alpha matting problem, we have used Equation 8 as given in Yan et al. [8], to calculate the gradient mixing parameter. We have used the alpha matting with pixel correlation to find the best sample for the image pixel in unknown region. The best sample are obtained using PSO algorithm. So, by using pixel correlation it is possible to find the relation of the pixel colour with respect to the neighbour, so that we do not lost the true pixel value in unknown region. For given image $I_s$ and $I_t$ as source and target images, to remove smudging artefacts from the cloned image alpha matting applied in image composition using alpha matting as given by Yan et al. [8].

$$ZI_c = \alpha_z I_s(p) + (1 - \alpha)I_t(p) \quad (8)$$

Where,
p is the inner pixel.
$I_s$ is the final modified composite image.
$\alpha_z$ is the unknown alpha matte.
$I_s(p)$ is the modified cloned region without smudging and discolour artefacts.

The value of $\alpha_z$ can be calculated from Equation 9 [8]:

$$\alpha_z = \frac{(I_z - I_t)(I_s - I_t)}{||I_s - I_t||^2} \quad (9)$$

Where,
$I_z$ is the observed colour of pixel z.
Assuming the given sample pair with the sample indexes f and b in candidate sample set of the foreground and the background image.

We have modified Equation 8 as in Equation 10, to calculate the alpha matte value based on image pixel correlation as the EMMVC Algorithm [1] has just used the constant value for the alpha. This modification is based on the pixel correlation for finding unknown region pixels value, as shown in Equation 10:

$$MZI_c = \alpha_z \quad (10)$$

Where,
$\alpha_z$ is the unknown alpha matte.
$MZI_c$ is the modified composite image using alpha matte.

To obtain the best value of unknown region we proposed this formula (Equation 11) that will find out the true pixel colour based on pixel correlation. So, Equation 3 has been modified by us to Equation 11:

$$MI_{hc}(p) = MZI_c * I_c(p) + (1 - \alpha)I_t(p) \quad (11)$$

Where,
$MI_{hc}$ is the modified pixel correlation based alpha matte for finding the best value for the unknown pixels.
p is the inner pixel.
$I_c$ is the cloned image pixel value.
$\alpha$ is the unknown alpha matte.
$I_t(p)$ is the target image pixel value.

Thus, Equation 1 has been enhanced to Equation 12 as a final Enhanced multilayer value cloning with illumination gradient mixing and image pixel correlation for alpha matting that can solve the limitations of the EMMVC method [1].

$$EI_{layers}(p) = \alpha * MHI_c(p) + \beta * MI_{hc}(p) \qquad (12)$$

Where,
$\alpha$ and $\beta$ are the blending parameters.
$MI_{hc}(p)$ is the modified pixel correlation based alpha matte for finding the best value.
p is the inner pixel.
$I_c$ is the cloned image pixel value.
$\alpha$ is the unknown alpha matte.
$I_t(p)$ is the target image pixel value.
$MHI_c(p)$ is the modified value for image discoloration and illumination difference calculation.

Table 2: Pseudocode for the proposed system.

---

**Algorithm: Enhanced Multi-layer mean value cloning Algorithm with illumination aware gradient mixing and image pixel correlation.**

**Input:** boundary pixels' list δ b, boundary inner pixels' list $b_{in}$, trimap with three regions source image $I_s$ and target image colour vector $I_t$.
**Output:** Merged video without artefacts.

**BEGIN**

**Step 1**: Generate the trimap for current and previous frames.
**Step 2**: If frame 0! = frame N then Calculate the contour flow between the two trimap using the trimap propagation contour flow method by the following Equation

$$I_{cf}(p) = (M_{xcurrent}, M_{ycurrent})$$

**Step 3**: Calculate the mean value coordinates of every inner pixel of generated source trimap generated in step 2 and the boundary pixels using the following Equation:

$$\lambda_i(p) = \frac{w_i}{\sum_{j=0}^{m-1} w_j} \quad i = 0,1, \ldots\ldots \text{ m-1}$$

**Step 4**: Repeat steps 2 and 3 to produce the cloned image for every image frame.
**Step 5**: Use the interpolation constraint coefficient to remove the discoloured artefacts in the cloned image here. k = 0 to 255, and it is considered as RGB channels. It can obtain by the following Equation:

$$MHI_c(p) = \begin{cases} Ma_i * I_s(p) + k * r(p) \text{ for frame 0} \\ Ma_i * I_{cf}(p) + k * Hr(p) \text{ for frames 1, 2 ... N} \end{cases}$$

Where,
$$Ma_i = \frac{||G_i^s||^2}{||G_i^s||^2 + ||G_i^t||^2}$$

*Note: We have added the illumination-aware gradient guided mixing algorithm step to remove any inconsistencies regarding the illumination light difference in the cloned image frame.*
**Step 6**: Alpha matte with image pixel correlation is applied to smooth and remove the smudging artefacts. It given by the following Equation:

$$MI_{hc}(p) = MZI_c * I_c(p) + (1 - \alpha)I_t(p)$$

**Step 7**: A multi layered merged video is produced by using volumetric image synthesisation to recover the occlusion. It can get by using the following Equation:

$$EI_{layers}(p) = \alpha * MHI_c(p) + \beta * MI_{hc}(p)$$

**END**

---

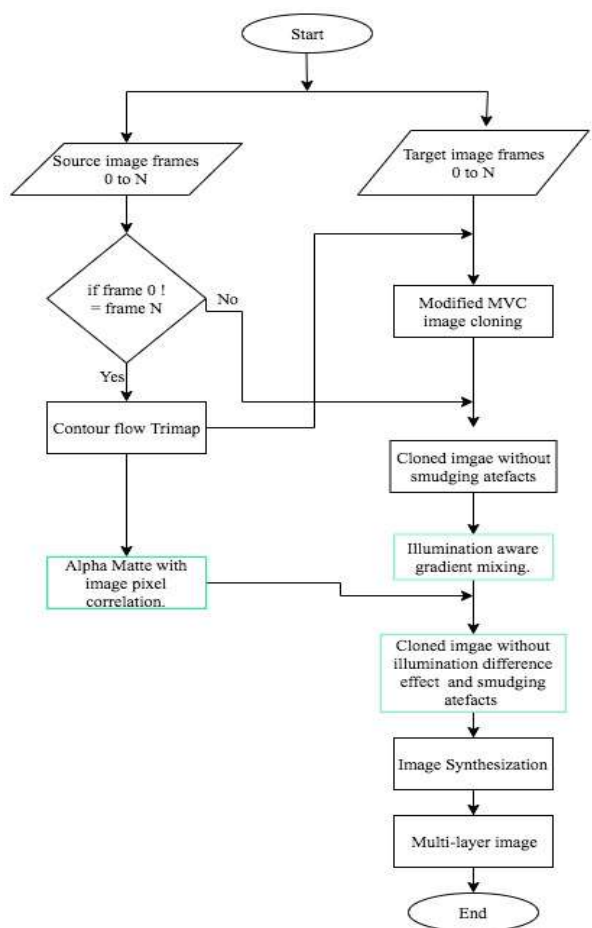

Figure 5: The Flowchart of the purposed system model.

## 3.2 Area of Improvement:

The main idea behind merging two videos is using EMMVC cloning with image interpolation and alpha matting. The proposed system uses the EMMVC and modify the existing alpha matting technique to produce a video with high accuracy and good consistency even in the illumination intensity varying inversely situation.

First, the illumination-aware gradient guided algorithm has been applied to address the problem mentioned in the limitation section of the EMMVC Algorithm [1]. We have added a new feature in the existing model that deals with the image frames whose illumination intensity varies with time. It uses gradient mixing parameter that is calculated from the gradient value of the source and the target image frames and then apply the mixing parameter to the cloned image frame to remove any inconsistencies in the final result.

To address the second issue, we have modified the pixel by pixel alpha matting approach with image pixel correlation which uses PSO algorithm to find the best sample value for a pixel in unknown region from the collected sample. The best pixel sample will be selected based on the relation with the neighbour pixels, i.e. other pixels in that unknown region. As a result, this will ensure that there is a very less chance of any artefacts like discolouration and smudging effect to occur in the final video.

Illumination aware grading mixing has been provided, so that the final merged video will have global temporal coherence even when the light intensity varies in the source and the target video using Equation 7. The EMMVC system [1] has not considered this issue. Second contribution is using the image pixel correlation to find the best sample in the sample collection for unknown pixel in unknown region when merging the image frames. and is given by Equation 11. The EMMVC [1] system has used the global alpha matting approach to find the unknown pixel and cannot find the best sample as they have not considered any relation between the neighbour pixels. All these solutions help to improve the accuracy and the visualisation of the final merged video.

# 4. Result and Discussion

Matlab R2019b have been used to implement the proposed system with 10 samples of breast (soft tissue) surgery videos, 10 samples of bowel (soft tissue) surgery video, and 10 samples of jaw (hard tissue) surgery videos. As a result, 30 different samples have been tested using 2.3 GHz Intel Core i5 16 GB memory and 8 GB graphic memory. These videos are collected from the online resources based on different age and weight groups. The length of the selected videos is from 10 seconds to 90 seconds. From those 10 videos, some frames are selected to be test using Matlab, and a hand video is recorded from which frames are extracted. The resolutions of the images used for testing are 920 by 545 pixels, 324 by 326 pixels and 310 by 326 pixels for breast, for breast, jaw, and bowel samples respectively. All the collected samples are tested for both the EMMVC system [1] and the proposed system and the results are shown in Table 3 and Table 4. The tested samples are divided into two types i.e. pre planning of breast surgery and intra operative of jaw surgery. The reason to select those two types is to show the clarity and the visualization of a hand on the patient body in the merging process. Visualization error also measured based on the RGB colour values in the resulted image of the EMMVC solution [1] and the proposed solution. We have also measured the MSE for all breast, bowel, and jaw samples.

At the beginning of the surgery, the expert surgeon needs to assist the rural surgeon. this can be done by the hand instruction given through the MR merged video. Venkata et al. [1] used the EMMVC technique but still it could not solve the issue with the light intensity difference and also does not use the best mechanism for artefacts removal in unknown region. To overcome this, Wang et al. [7] implemented gradient mixing for the illumination effect. Yan et al. [8] implemented pixel correlation for finding the best pixel value for unknown region. It improves the overlay and the visualization accuracy in the resulted merged image. Pre-operative mixed view is presented in Figure 6 and Intra operative mixed view is presented in Figure 7. Figures 6(a), 7(a) and 8(a) are the patient image, which was extracted from the augmented video. While Figures 6(b), 7(b) and 8(b) are the expert surgeon hand captured by the camera and Figures 6(c), 7(c) and 8(c) are the MR view to guide the local surgeon throughout the surgery for the breast, bowel, and jaw samples respectively. The results of merging (a) and (b) in (c) for all images types are presented in Table 3.

These results are generated by following our proposed algorithm as shown in Figure 4, Figure 5, and Table 2. The first step is to generate the trimap for current and previous frames. In the second step, frame 0 and frame N are compared and if they are not equal the contour flow will be calculated using the trimap propagation contour flow method. In the third step, the mean value coordinates are calculated of every inner pixel of the generated source trimap which is generated from the second. Similarly calculate the mean value coordinates for the boundary pixel. The above two steps are iterative and producing the cloned image for every image frame. After producing the cloned image, the interpolation constraint coefficient is used to remove the discoloured artefacts from the cloned image. In this step, we have added the illumination-aware gradient guided mixing algorithm to remove any inconsistencies regarding the illumination light difference in the cloned image frame. In step six, Alpha matte with image pixel correlation is applied to smooth and remove the smudging artefacts. In the last step, a multi layered merged video is produced by using volumetric image synthesisation to recover the occlusion, this is done using the proposed Equation 12.

Table 3: Result of merged video; Figure 5: breast sample; Figure 6: jaw sample; Figure 7: bowel sample.

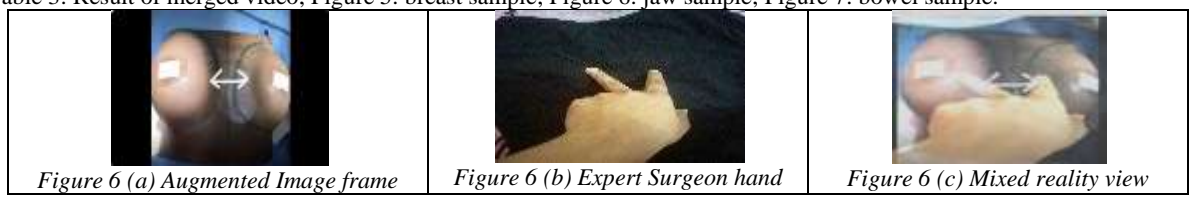

| Figure 6 (a) Augmented Image frame | Figure 6 (b) Expert Surgeon hand | Figure 6 (c) Mixed reality view |

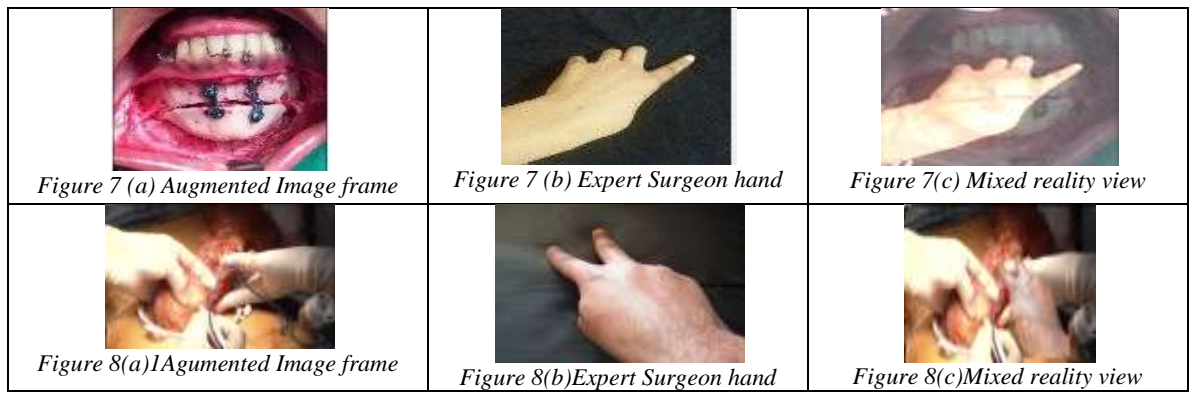

| Figure 7 (a) Augmented Image frame | Figure 7 (b) Expert Surgeon hand | Figure 7(c) Mixed reality view |
| Figure 8(a)1Agumented Image frame | Figure 8(b)Expert Surgeon hand | Figure 8(c)Mixed reality view |

Both soft and hard tissue samples are tested for the EMMVC Algorithm [1] and the proposed solutions using MATLAB and the results are presented in Tables 3, 4, 5 and 6. Each sample is measured and the average is calculated for accuracy of visualization error and the processing time, as shown in the three bar graphs of Figures 9, 10 and 11 respectively. The samples are divided into two parts i.e. pre-operative samples (soft tissue – breast, bowel) and intra operative samples (hard tissue – jaw). The accuracy for visualization error is calculated. Visualization error refers to the difference in RGB values between the original scene and the resulted scene. We have used imtool() function, a built in function in matlab. We passed our image frames as inputs to the function and then the image in a window is displayed which shows the pixel position and the respective RGB values of each pixel. The visibility of the pixel always depends on its RGB value. Finally, we collected the value of each pixel in the target area for both EMMVC Algorithm [1] and proposed system as shown in Tables 4-9. Processing time is taken as the average of all 10 samples and shown as a graph in Figures 9, 10, and 11. The tests have been done on 10 soft samples of breast and bowel and 10 hard tissue samples of jaw. The results of these tests are compared against the EMMVC solution [1] in term of visualization error, MSE, and processing time. The proposed system shows an improvement in the visualization of the hand on the patient body when there is illumination intensity difference between the body of patient and the expert instructor's hands. By using pixel correlation, we can also see the accurate RGB value in the unknown region when merging the results. Our solution reduces the visualization error and MSE compared to the EMMVC solution [1]. In case of processing time, there is no significant improvement.

Table 4: Accuracy of visualization error Results for Jaw samples: Results for accuracy of visualization error (Hard tissue)

| Sample number | Enhanced Multi-layer mean value cloning Algorithm [1] processed sample | Proposed system processed sample | Enhanced Multi-layer mean value cloning Algorithm[1] processed jaw sample RGB values | | | Proposed system processed Jaw sample RGB values | | |
|---|---|---|---|---|---|---|---|---|
| | | | R | G | B | R | G | B |
| 1 | 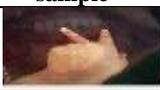 | 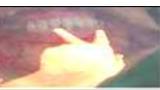 | 245 | 221 | 119 | 169 | 117 | 83 |
| 2 | 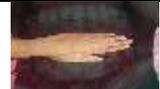 | 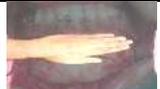 | 224 | 146 | 136 | 121 | 93 | 74 |
| 3 | 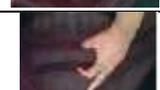 | 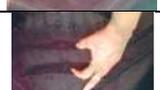 | 230 | 81 | 93 | 135 | 65 | 61 |
| 4 | 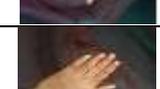 | 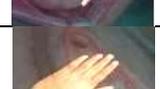 | 252 | 78 | 69 | 115 | 83 | 57 |
| 5 | 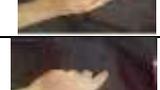 | 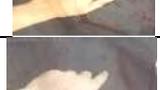 | 127 | 232 | 103 | 88 | 71 | 63 |
| 6 | 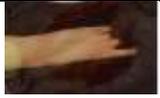 | 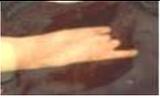 | 136 | 129 | 92 | 87 | 52 | 39 |
| 7 | 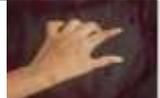 | 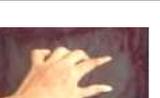 | 255 | 196 | 173 | 160 | 135 | 112 |
| 8 | 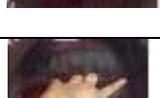 | 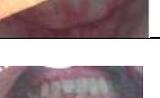 | 253 | 148 | 145 | 112 | 65 | 51 |
| 9 | 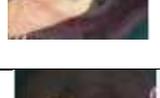 | 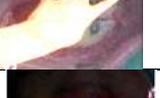 | 211 | 155 | 130 | 72 | 55 | 45 |
| 10 | 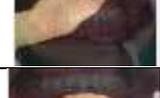 | 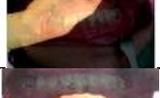 | 245 | 91 | 96 | 118 | 61 | 48 |
| **Average** | | | **196.2** | **129.6** | **134.4** | **144.6** | **93.4** | **83.33** |

Table 5 processing time for jaw samples for EMMVC Algorithm [1] and proposed system: Result table for processing time of Hard Tissue (Jaw):

| Sample Number | Sample details | AR Video (Patient images) | Expert hand | | Enhanced Multi-layer mean value cloning Algorithm [1] | | Proposed system | | |
|---|---|---|---|---|---|---|---|---|---|
| | | | | Processed Sample | Mean Squared Error (MSE) | Processing time/frame in seconds | Processed Sample | Mean Squared Error (MSE) | Processing time/frame in seconds |
| 1 | Lower front teeth | 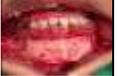 | 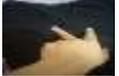 | 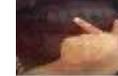 | **1.50** | **0.23** | 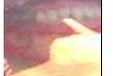 | **1.25** | **0.21** |
| 2 | Lower teeth | 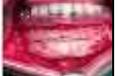 | 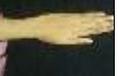 | 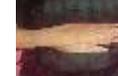 | **1.47** | **0.24** | 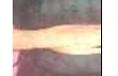 | **1.17** | **0.20** |
| 3 | Lower front gums | 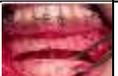 | 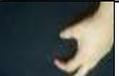 | 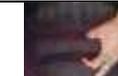 | **1.81** | **0.21** | 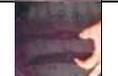 | **1.53** | **0.22** |
| 4 | Lower gingiva | 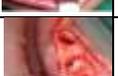 | 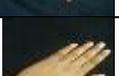 | 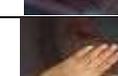 | 1.77 | 0.22 | 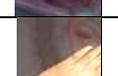 | **1.27** | **0.20** |
| 5 | Upper side teeth | 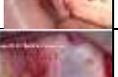 | 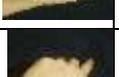 | 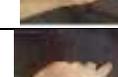 | **1.84** | **0.27** | 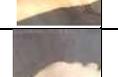 | **1.34** | **0.25** |
| 6 | Upper gingiva | 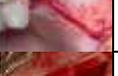 | 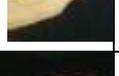 | 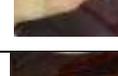 | **1.97** | **0.29** | 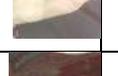 | **1.47** | **0.24** |
| 7 | Lower gingiva | 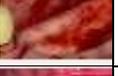 | 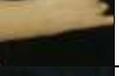 | 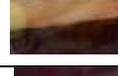 | **1.52** | **0.22** | 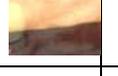 | **1.04** | **0.22** |
| 8 | Lower gingiva | 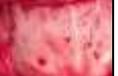 | 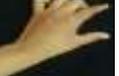 | 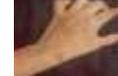 | **1.28** | **0.21** | 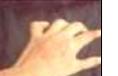 | **0.78** | **0.21** |

| 9 | Lower molar | 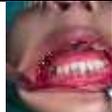 | 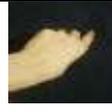 | 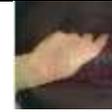 | 0.97 | 0.27 | 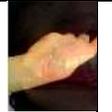 | 0.67 | 0.2 |
|---|---|---|---|---|---|---|---|---|---|
| 10 | Lower gingiva | 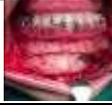 | 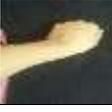 | 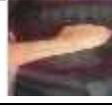 | 1.12 | 0.21 | 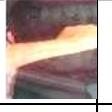 | 0.62 | 0.21 |
| Average | | | | | 13.78 | 0.202 | | 11.14 | 0.1934 |

Table 6: accuracy in term of visualization error for breast samples for EMMVC Algorithm [1] and proposed systems: Results for accuracy of visualization error (Soft tissue).

| Sample number | Enhanced Multi-layer mean value cloning Algorithm [1] processed sample | Proposed system processed sample | Enhanced Multi-layer mean value cloning Algorithm [1] processed breast sample RGB values | | | Proposed system processed breast sample RGB values | | |
|---|---|---|---|---|---|---|---|---|
| | | | R | G | B | R | G | B |
| 1 | 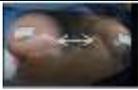 | 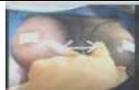 | 152 | 205 | 92 | 107 | 84 | 92 |
| 2 | 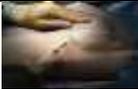 | 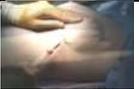 | 223 | 224 | 186 | 185 | 123 | 197 |
| 3 | 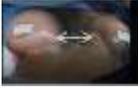 | 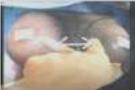 | 160 | 79 | 209 | 212 | 211 | 209 |
| 4 | 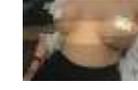 | 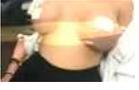 | 124 | 151 | 194 | 160 | 102 | 108 |
| 5 | 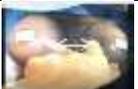 | 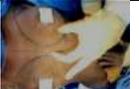 | 210 | 167 | 74 | 137 | 102 | 75 |
| 6 | 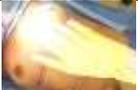 | 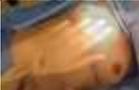 | 137 | 139 | 38 | 124 | 78 | 38 |
| 7 | 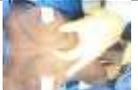 | 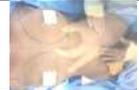 | 208 | 35 | 88 | 210 | 196 | 88 |
| 8 | 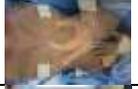 | 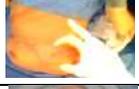 | 128 | 217 | 63 | 208 | 128 | 63 |
| 9 | 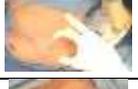 | 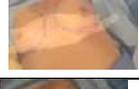 | 196 | 247 | 119 | 148 | 119 | 119 |
| 10 | 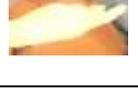 | 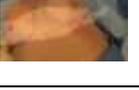 | 110 | 202 | 63 | 231 | 176 | 138 |
| Average | | | 149.3 | 173.9 | 114.3 | 142.7 | 113.2 | 91.5 |

Table 7: processing time for breast sample for EMMVC Algorithm [1] and proposed system: Result table time of Soft Tissue (Breast).

| Sample | Sample details (age, height weight (pounds)) | AR Video (Patient | Expert hand | | Enhanced Multi-layer mean value | Proposed system |
|---|---|---|---|---|---|---|

| | | images) | | | | cloning Algorithm[1] | | | | |
|---|---|---|---|---|---|---|---|---|---|---|
| | | | | | Processed Sample | Mean Squared Error (MSE) | Processing time/frame in seconds | Processed Sample | Mean Squared Error (MSE) | Processing time/frame in seconds |
| 1 | Breasts (31,5'6'',140) | 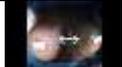 | 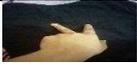 | | 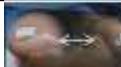 | 1.31 | 0.214 | 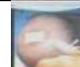 | 0.81 | 0.215 |
| 2 | Breasts (47,5'1'',115) | 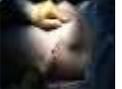 | 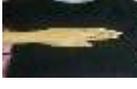 | | 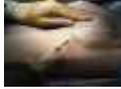 | 1.11 | 0.197 | 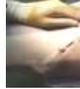 | 0.76 | 0.201 |
| 3 | Breasts (25,5'2'',120) | 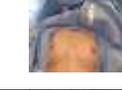 | 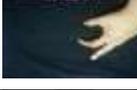 | | 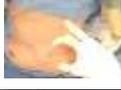 | 1.65 | 0.201 | 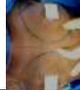 | 1.22 | 0.198 |
| 4 | Breasts (25,5'3'',160) | 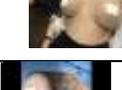 | 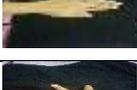 | | 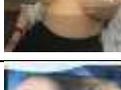 | 1.76 | 0.213 | 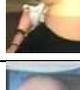 | 1.21 | 0.211 |
| 5 | Breasts (31,5'6'',140) | 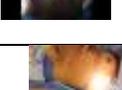 | 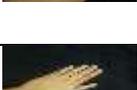 | | 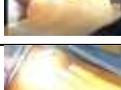 | 1.56 | 0.244 | 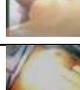 | 1.16 | 0.198 |
| 6 | Breasts (25,5'2'',120) | 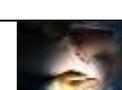 | 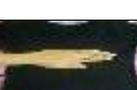 | | 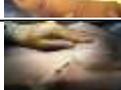 | 1.93 | 0.246 | 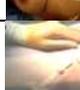 | 1.63 | 0.250 |
| 7 | Breasts (48,5'1'',115) | 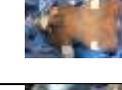 | 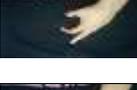 | | 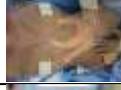 | 1.76 | 0.195 | 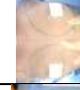 | 1.26 | 0.201 |
| 8 | Breasts (34,5'2'',190) | 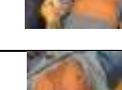 | 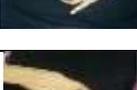 | | 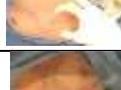 | 1.45 | 0.225 | 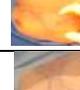 | 0.95 | 0.223 |
| 9 | Breasts (40,5'7'',155) | 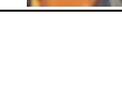 | 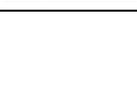 | | 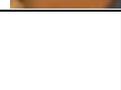 | 1.21 | 0.207 | 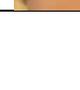 | 0.71 | 0.209 |
| 10 | Breasts (46,5'8'',142) | 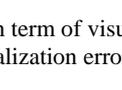 | 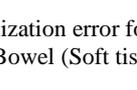 | | 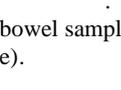 | 1.94 | 0.198 | 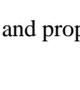 | 1.44 | 0.193 |
| Average | | | | | | 15.68 | 0.2094 | | 11.15 | 0.2069 |

.

Table 8: Accuracy in term of visualization error for bowel samples for EMMVC Algorithm [1] and proposed systems: Results for accuracy of visualization error Bowel (Soft tissue).

| Sample number | Enhanced Multi-layer mean value cloning Algorithm [1] processed sample | Proposed system processed sample | Enhanced Multi-layer mean value cloning Algorithm [1] processed bowel sample RGB values | | | Proposed system processed bowel sample RGB values | | |
|---|---|---|---|---|---|---|---|---|
| | | | R | G | B | R | G | B |
| 1 | 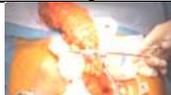 | 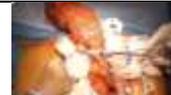 | 125 | 251 | 92 | 118 | 104 | 89 |
| 2 | 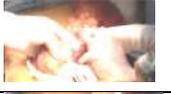 | 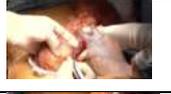 | 211 | 242 | 169 | 158 | 132 | 179 |
| 3 | 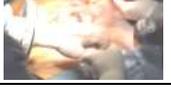 | 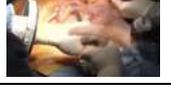 | 106 | 99 | 129 | 212 | 119 | 120 |
| 4 | 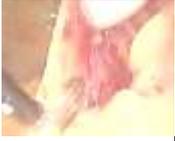 | 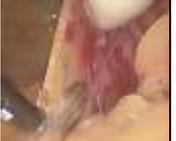 | 143 | 115 | 149 | 106 | 134 | 108 |
| 5 | 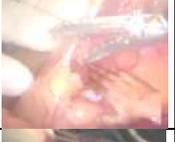 | 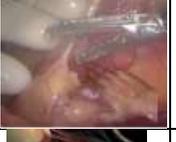 | 201 | 176 | 47 | 173 | 102 | 57 |
| 6 | 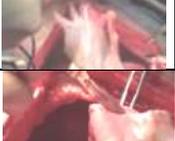 | 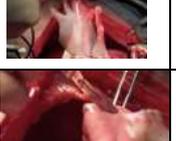 | 173 | 193 | 83 | 142 | 139 | 81 |
| 7 | 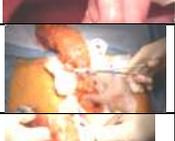 | 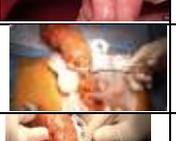 | 108 | 53 | 89 | 91 | 98 | 88 |
| 8 | 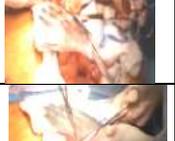 | 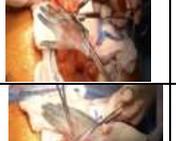 | 188 | 117 | 63 | 192 | 102 | 63 |
| 9 | 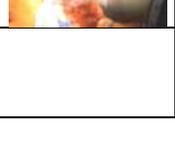 | 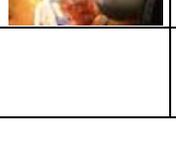 | 196 | 147 | 191 | 169 | 134 | 119 |
| 10 | 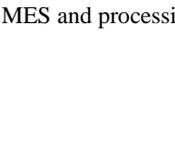 | 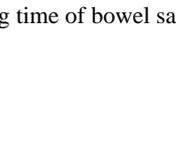 | 110 | 221 | 36 | 103 | 167 | 138 |
| **Average** | | | **156.1** | **161.4** | **104.8** | **146.4** | **123.1** | **104.2** |

Table 9: Result of MES and processing time of bowel sample: Result of MES and processing time of bowel sample.

| Sample Number | Sample details (age, height weight (pounds)) | AR Video (Patient images) | Expert hand | | Enhanced Multi-layer mean value cloning Algorithm[1] | | | Proposed system | |
|---|---|---|---|---|---|---|---|---|---|
| | | | | Processed Sample | Mean Squared Error (MSE) | Processing time/frame in seconds | Processed Sample | Mean Squared Error (MSE) | Processing time/frame in seconds |
| 1 | Bowel | 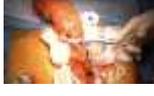 | 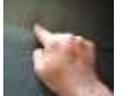 | 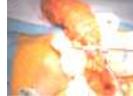 | 1.34 | 0.261 | 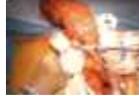 | 0.84 | 0.251 |
| 2 | Bowel | 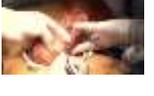 | 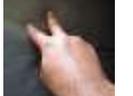 | 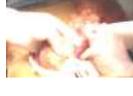 | 1.18 | 0.373 | 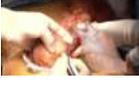 | 0.73 | 0.301 |
| 3 | Bowel | 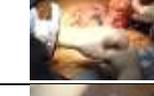 | 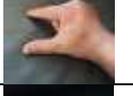 | 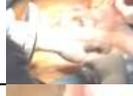 | 1.15 | 0.286 | 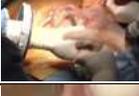 | 0.45 | 0.231 |
| 4 | Bowel | 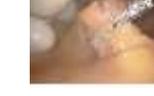 | 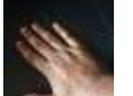 | 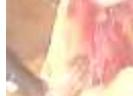 | 1.27 | 0.154 | 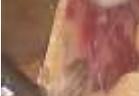 | 0.72 | 0.150 |
| 5 | Bowel | 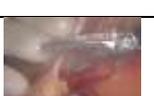 | 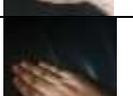 | 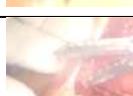 | 1.40 | 0.143 | 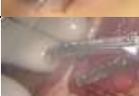 | 1.08 | 0.140 |
| 6 | Bowel | 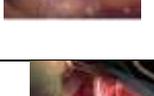 | 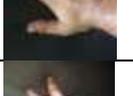 | 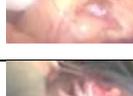 | 1.63 | 0.184 | 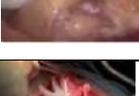 | 1.12 | 0.170 |
| 7 | Bowel | 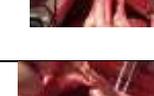 | 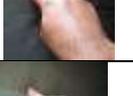 | 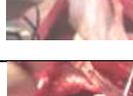 | 1.75 | 0.234 | 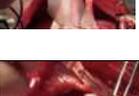 | 1.24 | 0.210 |
| 8 | Bowel | 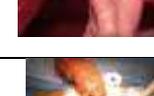 | 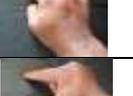 | 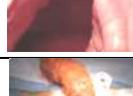 | 1.90 | 0.190 | 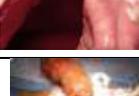 | 1.65 | 0.178 |
| 9 | Bowel | 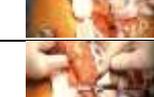 | 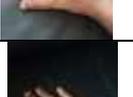 | 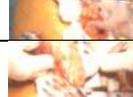 | 1.64 | 0.234 | 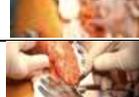 | 1.20 | 0.210 |
| 10 | Bowel | 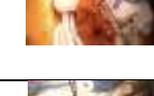 | 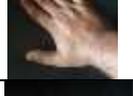 | 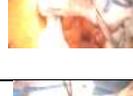 | 1.45 | 0.332 | 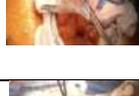 | 1.03 | 0.298 |
| Average | | | | | 14.71 | 2.391 | | 10.06 | 2.139 |

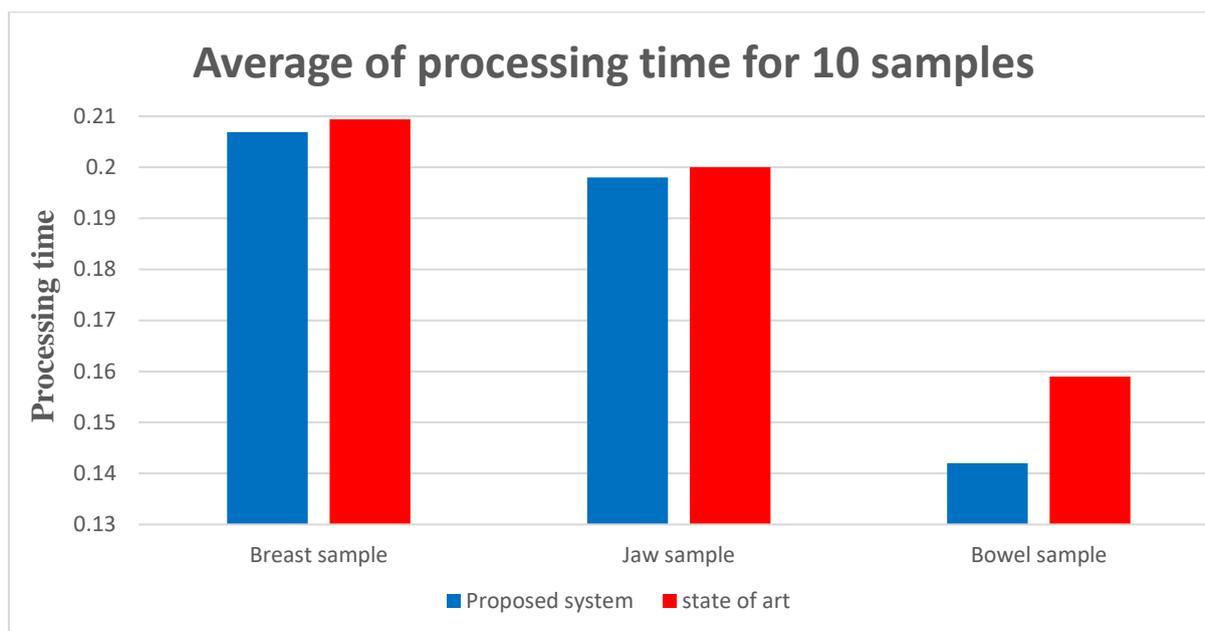

Figure 9: processing time in EMMVC Algorithm [1] and proposed solution for breast and jaw samples. a) first two bars show average processing time of EMMVC Algorithm [1] and proposed solution for breast samples. b) second two bar showing the average processing time of EMMVC Algorithm [1] and proposed solution for jaw samples.

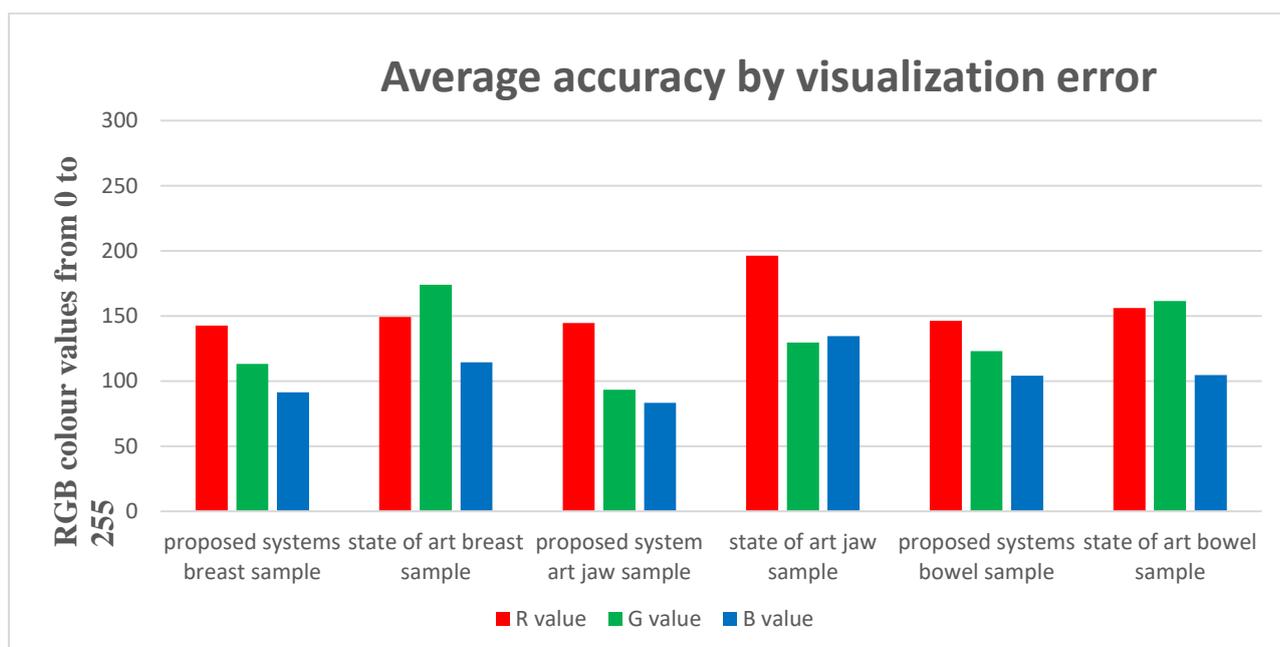

Figure 10: Average accuracy of visualization error for current and proposed solution with breast and jaw samples. a) First 3 bars showing the average of R value, G value and B value at particular pixel location of proposed solution with breast sample. b). Second 3 bars showing the average of R value, G value and B value at same pixel location of EMMVC solution [1] with breast sample. c) Third 3 bars showing the average of R value, G value and B value at particular pixel location of proposed solution with jaw sample. d) Last 3 bars showing the average of R value, G value and B value at same pixel location of EMMVC solution [1] with jaw sample.

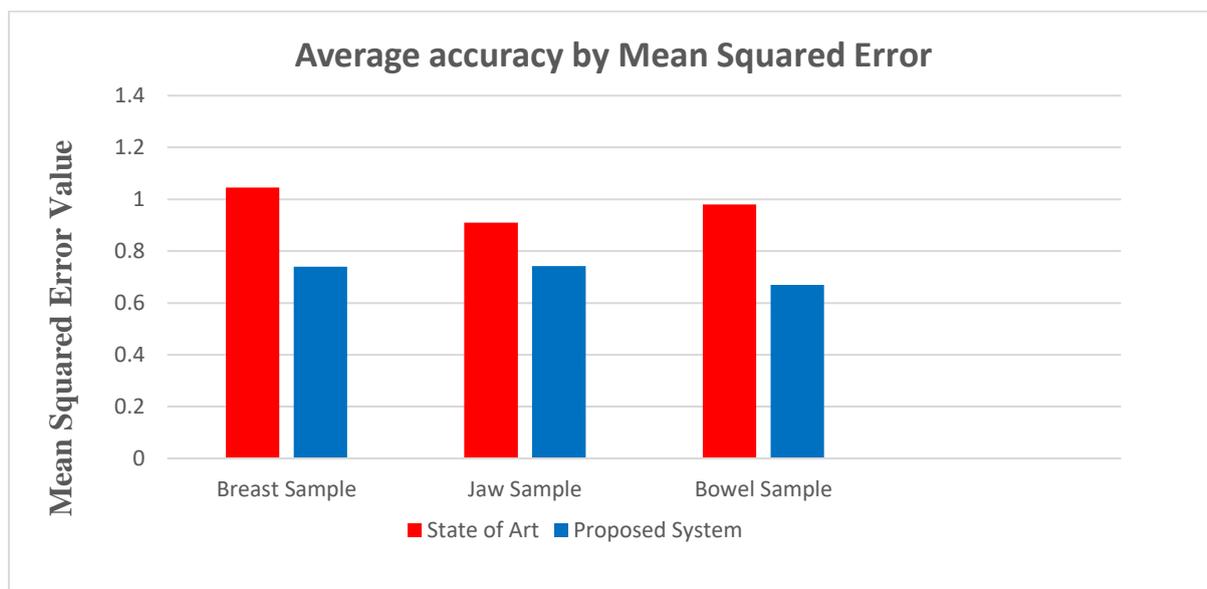

Figure 11: Average accuracy of Mean Squared error for current and proposed solution with breast, jaw and bowel samples.

The results presented in the Figures 9, 10, and 11, and the Tables 3-9 are a comparison between the EMMVC Algorithm [1] and the proposed solutions during merging. The proposed system shows an improvement in the visualization of hand in the light intensity difference situation. The proposed system also shows the true value of pixel in unknown region in the final merged image frame. We have compared the RGB value for the visualization error and the MSE error in the unknown region. From the above results, we can observe that there is a visible difference between the EMMVC system [1] and the proposed system in terms of the MSE and the visualization error. In the EMMVC Algorithm [1], no consideration has been taken for the light difference in the target and source image frame. while our proposed solution used illumination gradient mixing method to remove the illumination intensity difference. Furthermore, the proposed system can find the best value for unknown pixel in unknown region. We used the pixel correlation to find the best pixel by considering relation between the neighbours' pixels. The best sample values are collected using PSO algorithm. There is no big difference in the processing time for the two systems.

These measurements are calculated using MATLAB image tool. The visualization error is measured by converting the final image to RGB colour vectors and shown in Tables 3, 5 and 7. The difference between the EMMVC Algorithm [1] and the proposed solution can be observed by the RGB values. The EMMVC solution [1] suffers from highlighting problems and the proposed solution does not influence any effects. RGB values are ranged from 0 to 255 ('Zero' means black and 255 means white). In the EMMVC solution [1], some pixel values are measured nearly to 255 which represent a pixel under light effects. As a result, the hand could not visible clearly. In the same way, in intraoperative surgery, the internal body parts, and the blood flow can make the hand invisible. Values are measured for: R ranges from 50 to 255, G ranges from 80 to 255, and B ranges from 60 to 255. So, the proposed system can modularize the colour variations and the results showing that R value ranges from 70 to 210, G value ranges from 50 to 200 and B value ranges from 38 to 200,  assuming that  122 is the average pixel value for the range 0 to 255. For the tested samples, 99.1 % of the pixels are visualized by the current system and 99.7 % of the pixels are visualized by the proposed system. Anyway, it varies based on the image sample. Similarly, we calculated the MSE to check the best sample value for a pixel in an unknown region. The MSE value for the EMMVC Algorithm [1] and the proposed system shows that the proposed system is very close to finding the best pixel value as the MSE value is minimized compared to EMMVC system [1]. We have calculated the MSE value using the immse(), a build in function of MATLAB. We can see the average value for the 10 samples for jaw, breast and bowel is reduced from 13.78 to 11.14, from 15.68 to 11.15 and from 14.71 to 10.06 respectively. The illumination gradient mixing parameter helped to produce the globally consistent composite image frame where there is illumination intensity difference. In the same way it gave alpha matte value using image pixel correlation for the hand image in unknown region. Table 10 shows the comparison between the EMMVC Algorithm [1] and the proposed solution. The comparison is based on the accuracy, the processing time.

Table 10: Comparison Table between the proposed and the Enhanced Multi-layer mean value cloning solutions.

| | Proposed Solution | Enhanced Multi-layer mean value cloning Solution [1] |
|---|---|---|
| Name of the solution | Illumination-aware gradient mixing with Enhanced multi-layer mean value cloning | Enhanced multi-layer mean value cloning |
| Accuracy | The MSE has been improved by 16.38% as compared to EMMVC system. The visualization accuracy has been improved up to 99.7%. | The MES was only 62.73%. The visualization accuracy has been improved up to 99.1%. |
| Processing time | 50 frames in 9.7 seconds | 50 frames in 10 seconds |
| Contribution 1 | Illumination aware grading mixing algorithm has been used so that the final merged video will have global temporal coherence even when the light intensity varies in source and target video. | The EMMVC system has not considered any effect of light in source and target image frame when merging the video.. |
| Contribution 2 | Second contribution is using image pixel correlation to find the best sample in the sample collection for unknown pixel in unknown region when merging the image frames. | The EMMVC Algorithm has used the global alpha matting approach to find the unknown pixel and cannot find the best sample as they have not considered any relation between the neighbour pixels. |

## 5. Conclusion and Future Work

The augmented video generated at the local site is combined with the surgeon's hand's virtual video to produce a final video. The goal of this work is to produce high level of accuracy mixed reality view and send it to both sites. This combined video is capable of restoring the patient's augmented image pixels (which are obstructed by the hands and instruments of the surgeon) and virtual image pixels (which are influenced by the illumination intensity difference in operation theatre). The illumination aware grading mixing using Illumination Aware Video Composition algorithm has been provided so that the final merged video will have global temporal coherence even when the light intensity varies in source and target video. In addition, we have used an image pixel correlation to find the best sample in the sample collection for unknown pixel in unknown region when merging the image frames. The proposed solution helps to merge two videos by removing the effect of invers illumination intensity and reduce the error and smudging effect to improve visualization accuracy of the merged video and produce a globally consistent composite video. Our solution reduced the MSE rating to 16.88%, and the visibility improved from 99.1% to 99.7% even in inverse illumination intensity. Throughout the operation, this type of video helps the local surgeons to follow the guideline provided by the expert surgeon. Selecting the region of interest where the target is going to be merged in this system is a manual and time-consuming process. Future research should focus on automatic trimap generation so that the user does not have to identify the region of interest before merging the image frames. This will highly improve the performance and the processing time as there will be less human interaction and low possibilities to make errors. Also, furth research can be done to remove the noise caused by the blood flood during surgery.

**Abbreviation:**

| S.N. | Abbreviation | Full Form |
|---|---|---|
| 1 | AR | Augmented Reality |
| 2 | MR | Mixed Reality |
| 3 | MES | Mean Square Error |
| 4 | EMMVC | Enhanced Multi-layer Mean Value Cloning |
| 5 | PSO | Particle Swarm Algorithm |
| 6 | CT | Computed Tomography |
| 7 | AINET | Artificial Immune Neural Network |
| | | |

**References:**

[1] Venkata H. S. *et al.*, "A novel mixed reality in breast and constructive jaw surgical tele-presence," *Comput Methods Programs Biomed,* vol. 177, pp. 253-268, Aug 2019, doi: 10.1016/j.cmpb.2019.05.025.

[2] Shakya K., S. Khanal, A. Alsadoon, A. Elchouemi, *et al.*, "Remote Surgeon Hand Motion and Occlusion Removal in Mixed Reality in Breast Surgical Telepresence: Rural and Remote Care," *American Journal of Applied Sciences,* vol. 15, no. 11, pp. 497-509, 2018, doi: 10.3844/ajassp.2018.497.509.

[3] Fang, D., Xu, H., Yang, X. et al. An Augmented Reality-Based Method for Remote Collaborative Real-Time Assistance: from a System Perspective. Mobile Netw Appl 25, 412–425 (2020). https://doi.org/10.1007/s11036-019-01244-4.

[4] Cho D., Kim S., Tai Y.-W., and Kweon I. S, "Automatic Trimap Generation and Consistent Matting for Light-Field Images," *IEEE Transactions on Pattern Analysis and Machine Intelligence,* vol. 39, no. 8, pp. 1504-1517, 2017, doi: 10.1109/tpami.2016.2606397.

[5] Si W., Liao X., Qian Y., and Q. Wang Q., "Mixed Reality Guided Radiofrequency Needle Placement: A Pilot Study," in IEEE Access, vol. 6, pp. 31493-31502, 2018, doi: 10.1109/ACCESS.2018.2843378.

[6] Shen Y., Lin X., Gao Y., Sheng B., and Liu Q., "Video composition by optimized 3D mean-value coordinates," *Computer Animation and Virtual Worlds,* vol. 23, no. 3-4, pp. 179-190, 2012, doi: 10.1002/cav.1465.

[7] Wang J., Sheng B., Li P., Jin Y., and Feng D., "Illumination-Guided Video Composition via Gradient Consistency Optimization," *IEEE Trans Image Process,* May 20 2019, doi: 10.1109/TIP.2019.2916769.

[8] Yan X., Hao Z., and Huang H., "Alpha matting with image pixel correlation," *International Journal of Machine Learning and Cybernetics,* vol. 9, no. 4, pp. 621-627, 2019, doi: 10.1007/s13042-016-0584-1.

[9] Pluhacek M., Senkerik R., and Zelinka I., "Particle swarm optimization algorithm driven by multichaotic number generator," *Soft Computing,* vol. 18, no. 4, pp. 631-639, 2014, doi: 10.1007/s00500-014-1222-z.

[10] Henry C. and Lee S.-W, "Automatic trimap generation and artifact reduction in alpha matte using unknown region detection," *Expert Systems with Applications,* vol. 133, pp. 242-259, 2019, doi: 10.1016/j.eswa.2019.05.019.

[11] Archer J., Leach G., and van-Schyndel R., "GPU based techniques for deep image merging," *Computational Visual Media,* vol. 4, no. 3, pp. 277-285, 2018, doi: 10.1007/s41095-018-0118-8.

[12] Huang W., Alem L., Tecchia F., and Duh H., "Augmented 3D hands: a gesture-based mixed reality system for distributed collaboration," *Journal on Multimodal User Interfaces,* vol. 12, no. 2, pp. 77-89, 2017, doi: 10.1007/s12193-017-0250-2.

[13] Wang P. *et al.*, "2.5DHANDS: a gesture-based MR remote collaborative platform," *The International Journal of Advanced Manufacturing Technology,* vol. 102, no. 5-8, pp. 1339-1353, 2019, doi: 10.1007/s00170-018-03237-1.

[14] De Lima E. S., Feijó B., and Furtado A. L., "Video-based interactive storytelling using real-time video compositing techniques," *Multimedia Tools and Applications,* vol. 77, no. 2, pp. 2333-2357, 2017, doi: 10.1007/s11042-017-4423-5.

[15] Wang H., Xu N., Raskar R., and Ahuja N., "Videoshop: A new framework for spatio-temporal video editing in gradient domain," *Graphical Models,* vol. 69, no. 1, pp. 57-70, 2007, doi: 10.1016/j.gmod.2006.06.002.

[16] Basnet B. R., Alsadoon A., Withana C., Deva C. A., and Paul M., "A novel noise filtered and occlusion removal: navigational accuracy in augmented reality-based constructive jaw surgery," *Oral Maxillofac Surg,* vol. 22, no. 4, pp. 385-401, Dec 2018, doi: 10.1007/s10006-018-0719-5.

[17] Murugesan Y. P., Alsadoon A., Manoranjan P., and Prasad P. W. C., "A novel rotational matrix and translation vector algorithm: geometric accuracy for augmented reality in oral and maxillofacial surgeries," *Int J Med Robot,* vol. 14, no. 3, p. e1889, Jun 2018, doi: 10.1002/rcs.1889.

[18] Zhang M., Piao Y., Wei C., and Si Z., "Occlusion removal based on epipolar plane images in integral imaging system," *Optics & Laser Technology,* vol. 120, 2019, doi: 10.1016/j.optlastec.2019.105680.

[19] Oyekan J., Prabhu V., Tiwari A., Baskaran V., Burgess M., and McNally R., "Remote real-time collaboration through synchronous exchange of digitised human–workpiece interactions," *Future Generation Computer Systems,* vol. 67, pp. 83-93, 2017, doi: 10.1016/j.future.2016.08.012.

[20] Choi S. H., Kim M., and Lee J. Y., "Situation-dependent remote AR collaborations: Image-based collaboration using a 3D perspective map and live video-based collaboration with a synchronized VR mode," *Computers in Industry,* vol. 101, pp. 51-66, 2018, doi: 10.1016/j.compind.2018.06.006.